\documentclass[10pt,twocolumn,letterpaper]{article}

\usepackage[pagenumbers]{cvpr} % To force page numbers, e.g. for an arXiv version

\usepackage{graphicx}
\usepackage{amsmath}
\usepackage{amssymb}
\usepackage{booktabs}
\usepackage{multicol} % 用于多栏相关设置
\usepackage{caption} % 更好地控制图片标题等格式

\usepackage[pagebackref,breaklinks,colorlinks]{hyperref}

\usepackage[capitalize]{cleveref}
\crefname{section}{Sec.}{Secs.}
\Crefname{section}{Section}{Sections}
\Crefname{table}{Table}{Tables}
\crefname{table}{Tab.}{Tabs.}

\def\cvprPaperID{*****} % *** Enter the CVPR Paper ID here
\def\confName{CVPR}
\def\confYear{2022}

\begin{document}

\title{Boosting the Accuracy of Stock Market Prediction via Multi-Layer Hybrid MTL Structure}

\author{Yuxi Hong\\
School of Economics and Management, Harbin Institute of Technology\\
Weihai 264209, China\\
{\tt\small 2022210917@stu.hit.edu.cn}}
\maketitle

%%%%%%%%% ABSTRACT
\begin{abstract}
   Accurate stock market prediction provides great opportunities for informed decision-making, yet existing methods struggle with financial data's non-linear, high-dimensional, and volatile characteristics. Advanced predictive models are needed to effectively address these complexities. This paper proposes a novel multi-layer hybrid multi-task learning (MTL) framework aimed at achieving more efficient stock market predictions. It involves a Transformer encoder to extract complex correspondences between various input features, a Bidirectional Gated Recurrent Unit (BiGRU) to capture long-term temporal relationships, and a Kolmogorov-Arnold Network (KAN) to enhance the learning process. Experimental evaluations indicate that the proposed learning structure achieves great performance, with an MAE as low as 1.078, a MAPE as low as 0.012, and an  $R^2$ as high as 0.98, when compared with other competitive networks.
\end{abstract}

%%%%%%%%% BODY TEXT
\section{Introduction}
\label{sec:intro}
Accurate stock market prediction has long been a central focus in financial research \cite{01}. For investors, precise forecasts provide critical insights for investment strategies, while for corporations, they enable effective capital allocation and strategic planning. Stock price data, characterized as complex, non-stationary, and non-linear time series with substantial noise, present a significant challenge for predictive modeling \cite{02}.

Early methods, including ARMA \cite{3}, ARIMA \cite{4}, and GARCH \cite{5}, primarily relied on historical price data and traditional statistical frameworks. However, these models exhibit limited performance with non-linear, non-stationary financial data and often require manual hyperparameter tuning, leading to suboptimal real-world predictions \cite{6}. With the rise of artificial intelligence, the development of prediction models has shifted from basic statistical methods to complex machine learning techniques. Verified by scholars, the accuracy of machine learning technology to predict stocks is significantly better than traditional ones \cite{7}.

With the advancement of computing power and data analysis methods, deep learning technology has deepened the number of hidden layers on the basis of neural networks. Multi-layer artificial neural networks can be used to approximate any nonlinear function. In stock market data prediction, it shows stronger learning and generalization ability than traditional machine learning \cite{8}. The convolution neural networks \cite{9}, recurrent neural network \cite{10} and varient models \cite{11} have gradually received extensive attention in stock market price and return forecasting.

Nonetheless, conventional stock forecasting techniques primarily analyze individual stock time series in isolation, thereby overlooking the intricate relationships between different stocks.  Empirical evidence suggests that integrating the ’spatial’ dimension, which encompasses the relational information among stocks, can significantly enhance the predictive accuracy of models \cite{12}.

To achieve more accurate stock market predictions, it is crucial to consider the relationships between stocks. Indeed, a limited number of studies have employed graph neural network (GNN) methods to learn the impact of pairwise relationships on stock price prediction \cite{13}. However, in the real market, the relationships between different stocks are often more complex than simple pairwise interactions.

Transformer models, leveraging their attention mechanism, excel at capturing complex relationships across multi-source data, making them ideal for stock prediction by understanding long-term dependencies and multidimensional features \cite{14}.
The self-attention mechanism is computationally expensive, especially for large datasets or long time series. Moreover, Transformers lack inductive bias for time-series data, making them less effective at capturing seasonality and trends inherent in financial data. Additionally, they struggle with very long sequences and high-dimensional data. \cite{15}. 

To address these limitations, KAN (Knowledge-Augmented Network) replaces traditional weight parameters with edge function parameters, improving the approximation of nonlinear functions. Its fine-grained node adjustments enable better handling of complex, high-dimensional stock data, significantly enhancing prediction accuracy.

Ensemble learning is a crucial method for improving prediction accuracy \cite{16,17,18}. In the field of financial forecasting, ensemble learning has demonstrated increasingly significant advantages \cite{19,20}. Our algorithm can serve as a component within an ensemble learning framework to further enhance prediction precision.

Based on the above, we propose a multi-layer hybrid multi-task learning structure. The main contributions of our work are listed as follows:

\begin{itemize}
\item The paper introduces a new multi-layer hybrid MTL framework designed to achieve more efficient stock market predictions by addressing the non-linear, high-dimensional, and volatile characteristics of financial data. It combines the strengths of different layers, including a Transformer encoder, BiGRU, and KAN, to enhance overall predictive performance. 
\item The method can simultaneously extract the coupling relationship between different channel data and capture the strong correspondence in time series, so as to achieve joint optimization of all prediction tasks.
\item The framework uses a Transformer encoder to extract 
complex correspondences among various input features, enabling the model to capture deeper associations within multidimensional data. It employs a BiGRU to capture long-term temporal relationships and a KAN to augment the learning process, resulting in superior performance at the metrics of MAE, MAPE and $R^2$.
\end{itemize}

%-------------------------------------------------------------------------

%-------------------------------------------------------------------------
% \begin{equation}
% f(A)
% A=\{(x_1,...,x_m)|m=1,2...\}
% x_m=(a_{m1},a_{m2},...,a_{mn})
% f(A)=(a_{mn}+1,a_{mn}+2,...,a_{mn}+T)
% Q=XW_Q,K=XW_K,V=XW_V
% Q_i=XW_Q^i,K_i=XW_K^i,V_i=XW_V^i
% A_i=softmax(\frac{Q_iK_i^T}{\sqrt{d_k}})
% Z=Concat(Z_1,Z_2,...,Z_h)W_O
% FFN=max(0,xW^1+b^1)W^2+b^2
% KAN(x)=(\Phi_{L-1}\odot\Phi_{L-2}\odot...\odot\Phi_{0})
% \vec{h}_{ft} =GRU(\vec{h}_{f(t-1)},x_t)
% \vec{h}_{bt} =GRU(\vec{h}_{b(t+1)},x_t)
% h_t=[\vec{h}_{ft};\vec{h}_{bt}]
% y_t=softmax(Wh_t+b)
% MAE=\frac{1}{m}\sum_{i=1}^m|y_i-\bar{y}_i|
% RMSE=\sqrt{MSE}
% MAPE=\frac{100%}{n}\sum_{i=1}^m|\frac{y_i-\bar{y}_i}{y_i}|    
%R^2=I-\frac{SS_{residual}}{SS_{total}}
% \end{equation}

% Update the cvpr.cls to do the following automatically.
% For this citation style, keep multiple citations in numerical (not
% chronological) order, so prefer \cite{Alpher03,Alpher02,Authors14} to
% \cite{Alpher02,Alpher03,Authors14}.

%------------------------------------------------------------------------

\section{Related Work}
\label{sec:formatting}
Predicting stock market trends has long been of significant interest to both academics and practitioners due to its profound financial implications. The literature on stock market forecasting encompasses a wide array of methods, ranging from traditional statistical approaches to advanced machine learning models and hybrid systems that combine multiple techniques. Traditional methods have historically dominated the field. However, with the advent of sophisticated machine learning algorithms, including neural networks and deep learning architectures, there has been a shift towards more complex and adaptive predictive models. Furthermore, hybrid models that integrate the strengths of both traditional and modern techniques have emerged as promising solutions, offering enhanced accuracy and robustness in dynamic environments.
%-------------------------------------------------------------------------
\subsection{Traditional Methods}
Traditional methods for addressing the problem of stock market forecasting can be categorized into traditional time series models and hidden Markov models. Regarding traditional time series models, Devi et al.\cite{21} collected stock data spanning the past five years and, for the first time, utilized ARIMA models with varying parameters to predict future market trends. Khanderwal et al.\cite{22}noted that the ARIMA model holds potential for satisfactory short-term stock price prediction. Marisetty et al. \cite{23} employed a series of GARCH models to analyze the volatility dynamics of five major global financial indices over a decade, concluding that the GARCH model is the preferred choice for volatility prediction. Among hidden Markov models, a notable contribution is the work of Gupta et al. \cite{24} who proposed a Maximum A Posteriori (MAP) estimator based on Hidden Markov Models (HMM) to predict the next day’s stock values using historical data. Their study concluded that this model exhibited higher performance compared to ARIMA and ANN models. Su et al. \cite{25}further extended the HMM framework by adapting it from a discrete to a continuous form, essentially proposing a model for predicting stock price trends. Since these traditional statistical methods are fundamentally linear, they may hinder prediction performance in challenge scenarios involving sudden fluctuations in stock prices. Recent research efforts have focused on optimizing these traditional methods. For instance, Mattera et al. \cite{26}introduced a dynamic network autoregressive conditional heteroskedasticity (ARCH) model, which greatly addresses the limitation that multivariate ARCH models are generally unsuitable for high-dimensional input data, thereby enhancing their total stock prediction accuracy.

\subsection{Machine Learning Networks}
In the current era of artificial intelligence, machine learning plays a relevant role in the prediction of time series. Through a comparative analysis of traditional methods and machine learning approaches, Bhattacharjee et al. \cite{27}concluded that machine learning methods, particularly neural network models, exhibit the highest accuracy in stock price prediction. Vijh et al. \cite{28} utilized artificial neural networks and random forest techniques to forecast the next day’s closing prices of five companies across different sectors. As a subset of machine learning, deep learning has emerged as a prominent area of research in stock prediction due to its superior handling of complex, nonlinear data. Mehtab et al. \cite{29} employed eight machine learning and four LSTM-based deep learning regression models to predict stock index values and their change patterns over a weekly period, demonstrating that LSTM-based single-input deep learning models outperform machine learning-based models. However, the aforementioned research primarily relies on single data inputs, which are insufficient to describe the interrelationships among multiple variables. Given the messy, dynamic nature of stock market data and its dependence on economic factors, Assaf et al. \cite{30} demonstrated that multivariate deep learning can enhance stock market volatility prediction compared to single-input models. To validate the assertion, Tang et al. \cite{31} proposed model based on a wavelet transform-based LSTM that incorporating multi-dimensional data inputs, achieving an accuracy rate of 72.19\%, which surpasses that of single-input LSTM models. Deep et al. \cite{32} introduced a novel multifactor analysis model for stock market forecasting that extends beyond the use of deep learning models alone. This model integrates technical analysis, fundamental analysis, machine learning, and sentiment analysis, and demonstrates superior performance compared to single-factor models.

%-------------------------------------------------------------------------
\subsection{Hybrid Models}
Given the inherent advantages and limitations of the techniques mentioned, some researchers have sought to enhance prediction accuracy through the development of hybrid learning models. Hybrid models, which are based on machine learning, combine the predictions of one or more methods to improve the accuracy of simple predictions and mitigate potential overfitting issues. The concept of ensemble models was first proposed in 2001. Abraham et al. \cite{33} constructed one of the earliest hybrid models based on neural fuzzy logic and artificial neural networks (ANN), confirming that hybrid systems yield promising results for both prediction and trend forecasting, warranting further research and analysis. Subsequently, Shah et al. \cite{34} discussed various forecasting methods that could aid in predicting stock prices, highlighting that a hybrid approach combining statistical and machine learning techniques might prove more effective for stock forecasting. Shui-Ling et al. \cite{35} proposed a novel hybrid ARIMA-RNN model to predict stock prices, addressing the volatility issue of single models and avoiding the overfitting problem associated with neural networks.

In recent years, research focus has shifted toward hybrid models incorporating deep learning. Zhang et al. \cite{36} introduced a stock index prediction model based on the ARIMA-CNN-LSTM architecture, demonstrating superior prediction accuracy and robustness compared to traditional models. Tian et al. \cite{37} introduced a multi-layer bidirectional LSTM-BO-LightGBM model to address stock price volatility prediction, verifying its enhanced approximation and generalization capabilities relative to single models. Lv et al. \cite{38} highlighted that there is still room for improving prediction accuracy by transitioning from single deep learning prediction models to hybrid models. They proposed a deep learning hybrid model, CEEMDAN-DAE-LSTM, for stock index prediction, introducing DAE as a feature extraction module for the first time. This module extracts more nonlinear features of the stock index, thereby enhancing the model's prediction capability for volatile stock indices. These studies indicate that ensemble learning models have become a prevalent approach for improving prediction accuracy, particularly when dealing with complex and highly dynamic datasets. Moreover, these methods emphasize the necessity of continuing to develop and refine diffrent hybrid models to address the challenges posed by changing market conditions and technological advancements.

%-------------------------------------------------------------------------
\section{The Proposed Method}
\subsection{Problem Statement}

In this paper, the objective of stock price prediction is to develop an effective mapping function $f(A)$, such that for input $A=\{(x_1,...,x_m)|m=1,2...\}$, where$x_m=(a_{m1},a_{m2},...,a_{mn})$,  the  function need to yield  a  set  of  predicted outputs$f(A)=(a_{mn}+1,a_{mn}+2,...,a_{mn}+T)$ , in which  denotes the steps we need to forecast. The goal is to ensure that the predicted outputs closely approximate the actual values.
\subsection{Overview of the Multi-Layer Hybrid MTL Structure}
The proposed structure integrates multiple dimensions of financial data to enhance predictive accuracy. The framework takes as input a variety of market indicators, including the opening price, closing price, lowest price, highest price, trading volume, and trading amount. These diverse inputs provide a comprehensive representation of the market dynamics, allowing the model to capture intricate relationships and patterns that are essential for accurate forecasting. Fig. 1 provides the proposed framework, which consists of several key components, including a Transformer encoder, a KAN layer, and a BiGRU layer. Table 1 illustrates the configuration of the structure, in which Transformer encoder is responsible for capturing the correspondences among various input features, enabling the model to handle high-dimensional data efficiently. KAN further refines the  learning process, enhancing the model's ability to extract meaningful representations from the input data. BiGRU component captures the long-term temporal dependencies present in financial time series data. By incorporating bidirectional information flow, the BiGRU ensures that both past and future contexts are considered simultaneously, thus providing a richer understanding of the temporal dynamics within the dataset. After processing through these layers, the model generates predictions for both trading volume and trading amount. This multi-task learning approach leverages the shared latent representations learned from the diverse input features to improve the overall stock predictive performance. By integrating these components, the proposed MTL structure aims to deliver robust and accurate forecasts, addressing the complexities inherent in financial markets.

\begin{table}[htbp]
\centering
\caption{Detailed configuration of the proposed structure.}
\begin{tabular}{lll}
\hline
\textbf{Layer(type)} & \textbf{Output Shape }& \textbf{Param\#} \\
\hline
Input Layer & [(None, 5, 1)] & 0 \\
Transformer & [(None, 5, 32)] & 4 \\
KAN & [(None, 5, 25, 50)] & 2 \\
GRU & [(None, 5, 32)] & 3360 \\
Bidirectional & [(None, 5, 256)] & 164864 \\
Batch Normalization & [(None, 5, 256)] & 1024 \\
Dropout & [(None, 5, 256)] & 0 \\
Permute & [(None, 256, 5)] & 0 \\
Dense & [(None, 256, 5)] & 30 \\
Permute & [(None, 5, 256)] & 0 \\
Multiply & [(None, 5, 256)] & 0 \\
Batch Normalization & [(None, 5, 256)] & 1024 \\
Dropout & [(None, 5, 256)] & 0 \\
Flatten & [(None, 1280)] & 0 \\
Dropout & [(None, 1280)] & 0 \\
Dense & [(None, 1)] & 1281 \\
\hline
\end{tabular}
\end{table}

%figure 

\begin{figure*}
  \centering
    %\fbox{\rule{0pt}{2in} \rule{.9\linewidth}{0pt}}
    \includegraphics[width=1\linewidth]{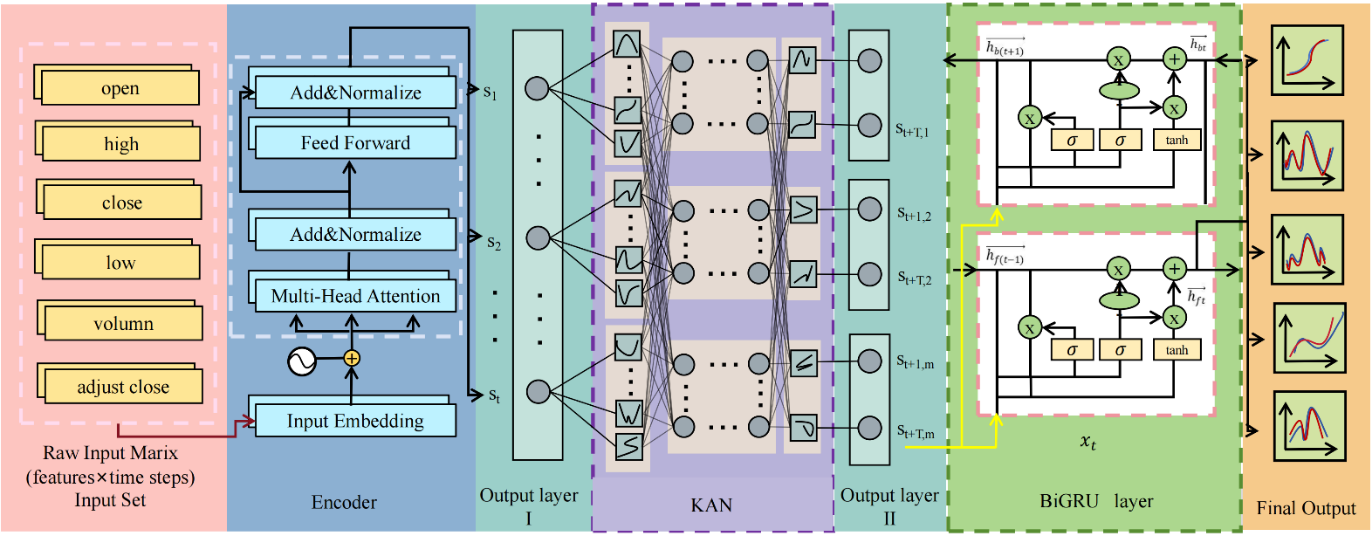}
    \label{fig:short-b}
  \caption{The multi-layer hybrid MTL structure for stock market prediction, involving an enhanced Transformer, KAN and BiGRU.}
  \label{fig:short}
\end{figure*}

%Table 1. Detailed configuration of the proposed structure.

%-------------------------------------------------------------------------
\subsection{Transformer Encoder Layer}

Each Transformer encoder layer comprises two sub-layers: a multi-head self-attention mechanism and a fully connected feed-forward network, both equipped with residual connections. Layer normalization operations are applied around each sub-layer. Fig. 2 illustrates the workflow of the attention mechanism, where the essence of the multi-head attention mechanism is the result of parallel computing of multiple independent attention mechanisms, and then integrated through the linear layer. Firstly, the input vector $X$ is transformed into three vectors: Query ($Q$), Key ($K$) and Value ($V$) by different linear transformations.
\begin{equation}
Q=XW_Q,K=XW_K,V=XW_V
\end{equation}

where $W_Q$,$W_K$ and $W_V$ are the weight matrices of different vectors respectively. These transformed matrices are then divided into the multiple heads. For each head $i$=1, 2,…,$h$, the corresponding query, key, value matrices are computed with,
\begin{equation}
Q_i=XW_Q^i,K_i=XW_K^i,V_i=XW_V^i
\end{equation}
Each head independently calculates its attention weights using the scaled dot-product attention mechanism, defined as:
\begin{equation}
A_i=softmax(\frac{Q_iK_i^T}{\sqrt{d_k}})
\end{equation}
where $d_k$ is the dimensionality of the key vectors. The attention weights$A_i$ are used to compute the weighted sum of the value vectors $V_i$, resulting in the output for the $i$-th head, which is  $Z_i=A-iV-I$. The outputs from all the heads are concatenated into a single matrix and passed through a linear transformation with a weight matrix $W_o$ to produce the output:
\begin{equation}
Z=Concat(Z_1,Z_2,...,Z_h)W_O
\end{equation}
We use 4 attention heads to achieve more accurate prediction accuracy. By using multiple attention heads, the model can focus on different feature spaces of the input data, thereby enhancing its ability to capture complex patterns and improving the overall performance.

\begin{figure}[t]
  \centering
   \includegraphics[width=1.1\linewidth]{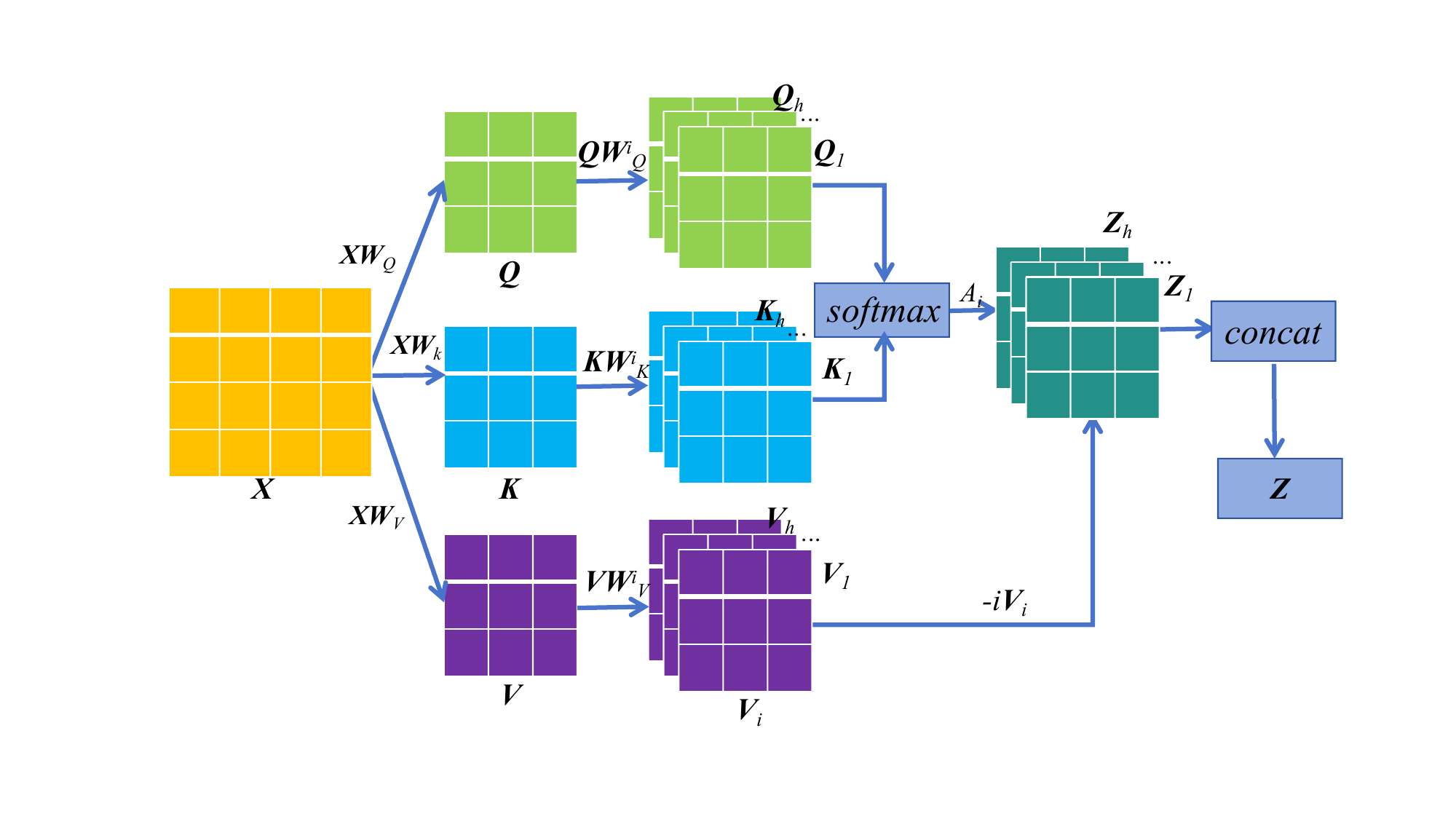}

   \caption{The multi-head attention in the Transformer structure.}
   \label{fig:onecol}
\end{figure}

The Feed Forward Layer is made up of a fully connected feed-forward network (FNN). In each encoder sub- module,FNN enhances the learning ability of the model through nonlinear transformation. The layer is composed of two linear transformations and a ReLU activation function, which greatly introduces nonlinear characteristics. This enables the model to capture more complex feature relationships and improve its performance in processing complex time series data of the stock market. The expression of this layer is as follows.
\begin{equation}
FFN=max(0,xW^1+b^1)W^2+b^2
\end{equation}
The Add/Normalize layer combines Layer Normalization and Residual Connection to reduce the problem of gradient disappearance or explosion. The output sequence of transformer will be transmitted as the input of KAN structure.

\subsection{Detailed KAN Layer}
Li et al. \cite{39} extended the use of Kolmogorov-Arnold theorem in machine learning and constructed a new KAN neural network architecture. Compared with MLP, the activation function of KAN is located on the edge rather than the node (see Fig. 3). In addition, MLP uses predefined activation functions (such as ReLU or sigmoid), while KAN allows learning adaptive nonlinear functions. At the same time, KAN can set fine-grained nodes to improve the approximation accuracy. 

The above characteristics make KAN have great advantages in dealing with smooth and nonlinear functions. After experimental verification, the curve of KAN can converge more quickly and reach a steady state, and it is better than the scaling curve of MLP, especially in the case of high dimension. It has been verified that it has improved prediction accuracy compared with traditional MLPs and is suitable for time series prediction \cite{40}. The architecture revolves around an innovative concept: traditional weight parameters are replaced by univariate function parameters at the edge of the network.  In KAN, each node summarizes the output of these functions without any nonlinear transformation. The KAN layer is 
\begin{equation}
 KAN(x)=(\Phi_{L-1}\odot\Phi_{L-2}\odot...\odot\Phi_{0})
\end{equation}
where $\Phi_{1}$ is the function matrix (B-spline function matrix) corresponding to the first KAN layer, and $x$ is the input matrix.
\begin{figure}[t]
  \centering
   \includegraphics[width=1\linewidth]{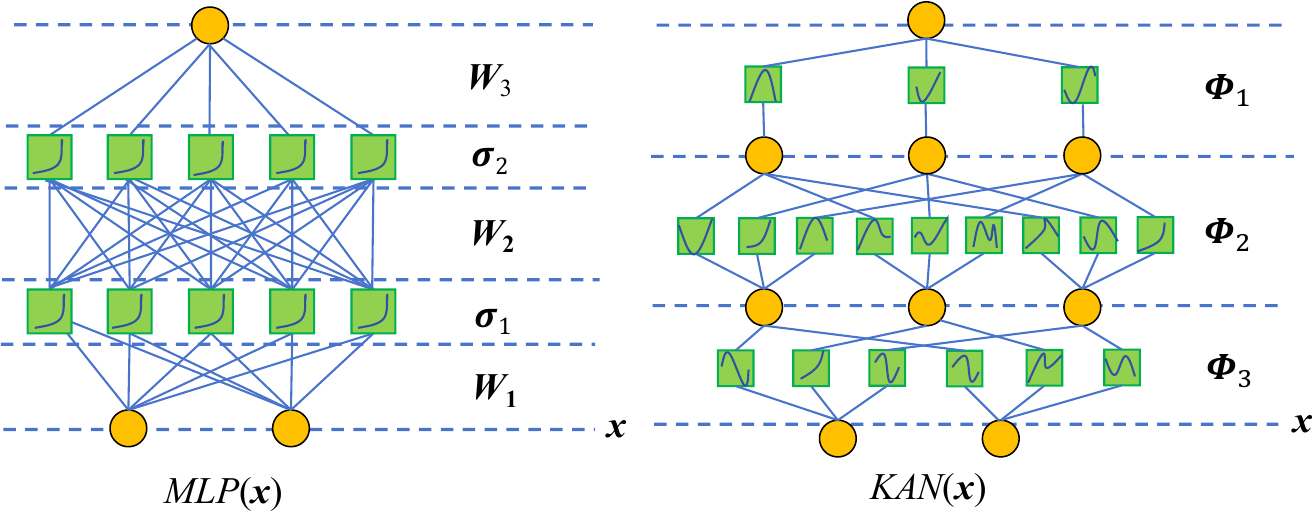}

   \caption{The detailed comparison of MLP and KAN structure}
   \label{fig:onecol}
\end{figure}

\subsection{Detailed BiGRU Layer}

Due to the complexity of the stock market and the volatility of the input data, this hybrid model adopts a two-way GRU structure.  BIGRU is a multivariate time series prediction method based on bidirectional GRU. It can adapt to more complex sequence patterns, and can effectively capture the two-way dependencies in time series data and the interaction between multiple variables. The BIGRU model is composed of GRU networks in two directions. One network processes time series data from forward to backward, and the other network processes time series data from backward to forward. That is,
\begin{equation}
 \vec{h}_{ft} =GRU(\vec{h}_{f(t-1)},x_t)
\end{equation}
\begin{equation}
\vec{h}_{bt} =GRU(\vec{h}_{b(t+1)},x_t)
\end{equation}
where$\vec{h}_{ft}$and $\vec{h}_{bt} $   represent the hidden state from left to right and from right to left respectively. GRU represents the GRU unit, and $x_t$ represents the $t$-th element in the input sequence. The final hidden state is the splicing of hidden states in two directions.
\begin{equation}
h_t=[\vec{h}_{ft};\vec{h}_{bt}]
\end{equation}
where [·; ·] denotes the splicing operation of vectors. Finally, the hidden state is passed to a fully connected layer to obtain the output $y_t$. The illustration of $y_t$ can be expressed as,
\begin{equation}
y_t=softmax(Wh_t+b)
\end{equation}
where $W$ and $b$ denote the weight and bias of the fully connected layer, respectively, and $softmax$ is the SoftMax activation function.
%-------------------------------------------------------------------------
\section{ EXPERIMENTAL EVALUATIONS}

\subsection{Experimental Design}

To rigorously evaluate and demonstrate the effectiveness of our proposed method, we conducted two types of experiments. We first compared our model against several representative advanced methods, including those proposed by Seo et al. \cite{41}, Ma et al. \cite{42}, Gao et al. \cite{43}, and Hemajothi et al. \cite{44}. Besides, we performed ablation experiments using five established models: KAN, Transformer, BiGRU, KAN-BiGRU, and Transformer-KAN. These comparisons were designed to provide a comprehensive assessment of the predictive performance and robustness of our proposed model in the context of stock price forecasting. The experiments involved multi-feature input and multi-objective prediction. The input features included the opening price (Open), closing price (Close), highest price (High), lowest price (Low), each of which impacts stock trading volume. In this study, the dataset contained all of the aforementioned input features, and the goal was to accurately predict and generate each target feature value across multiple future time steps. 
%-------------------------------------------------------------------------
\subsection{Evaluating Indicator}

To ensure a thorough evaluation, we employ four widely recognized metrics that collectively offer a multi-faceted view of model performance, including mean absolute error (MAE), root mean square error (RMSE), mean absolute percentage error (MAPE) and R-squared ($R^2$). If y represents the predicted value, $y_i$ represents the true value, and $\bar{y}_i$ is the number of samples, the representation of the four metrics are concluded as follows.

MAE evaluates the average absolute difference between predicted values and actual values of the model, which is very sensitive to outliers or large prediction errors. A smaller the MAE is, the stronger the prediction ability of the model is.
\begin{equation}
 MAE=\frac{1}{m}\sum_{i=1}^m|y_i-\bar{y}_i|
\end{equation}
     
RMSE restores the measurement of error to the same unit as the original data, making the results more intuitive and easier to understand. The representation is demonstrated as follow.
\begin{equation}
 RMSE=\sqrt{MSE}
\end{equation}

MAPE is a measure of the prediction accuracy of prediction methods in statistics. It provides a relative error measure that enables it to compare the performance of different models on data sets of different sizes. A lower MAPE indicates that the prediction accuracy of the model is higher.
\begin{equation}
 MAPE=\frac{100\%}{n}\sum_{i=1}^m|\frac{y_i-\bar{y}_i}{y_i}|
\end{equation}

 $R^2$measures the proportion of the variance that can be predicted by the independent variable, indicating that the model fits the input data well. It is illustrated as follow.
\begin{equation}
R^2=I-\frac{SS_{residual}}{SS_{total}}
\end{equation}
where $SS_{residual}$  is residual sum of squares, and that $SS_{total}$ is the total sum of the squares. The four metrics are collected when finishing all the prediction experiments.

%------------------------------------------------------------------------
\subsection{Result Evaluation and Analysis with Advanced Methods}

Fig. 4, Fig.5, Fig.6, Fig. 7 and Fig 8 provides five representative comparison results of the stock features with different comparative advanced methods. The comparative results show that the proposed method maintains a high consistency with the real value in most time periods, especially in the interval of 0-50 and 120-200 time steps, which shows lower error than other methods. In contrast, other methods often lag or deviate too much in these intervals, and the proposed method can more accurately capture the overall trend and all the change inflection points. In the prediction of peaks and troughs, the proposed method also indicates obvious advantages. At multiple key points such as 40-60 and 170-200 time steps, it can not only accurately identify local highs and lows, but also avoid deviations from other methods under noise interference. In addition, the model shows good robustness in the high-frequency fluctuation region, which can effectively filter out noise and make the prediction curve smoother and closer to the real value.
\begin{figure*}
  \centering
    %\fbox{\rule{0pt}{2in} \rule{.9\linewidth}{0pt}}
    \includegraphics[width=1\linewidth]{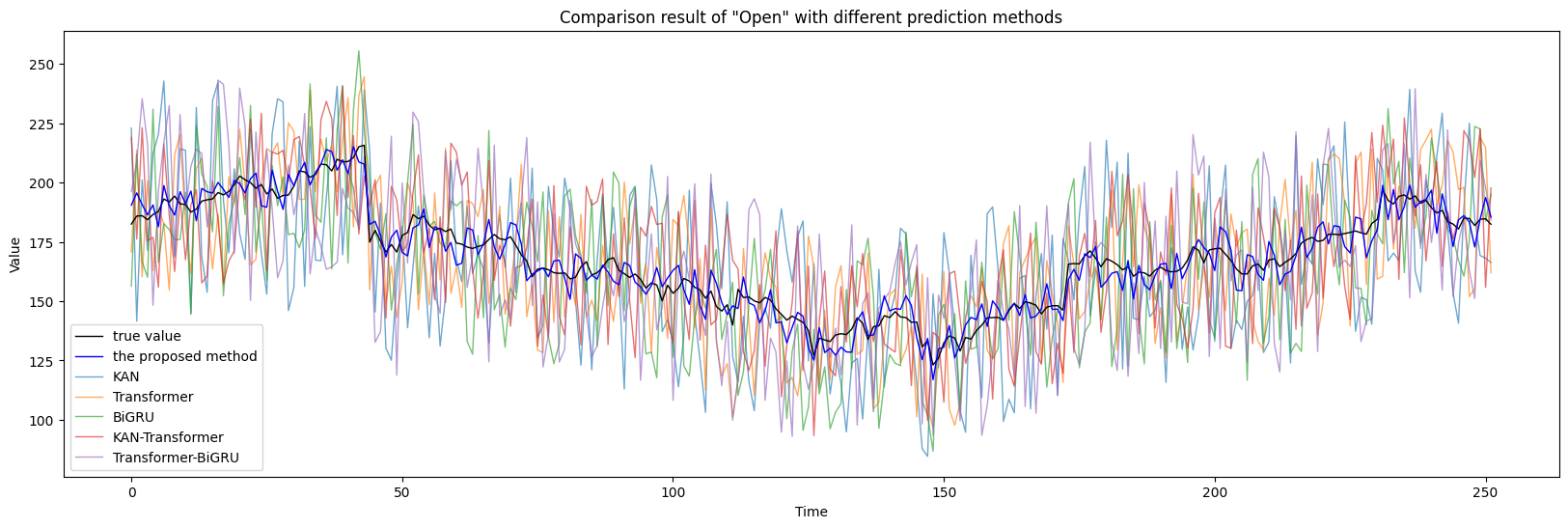}
    \label{fig:short-b}
  \caption{The multi-layer hybrid MTL structure for stock market prediction, involving an enhanced Transformer, KAN and BiGRU.}
  \label{fig:short}
\end{figure*}
\begin{figure*}
  \centering
    %\fbox{\rule{0pt}{2in} \rule{.9\linewidth}{0pt}}
    \includegraphics[width=1\linewidth]{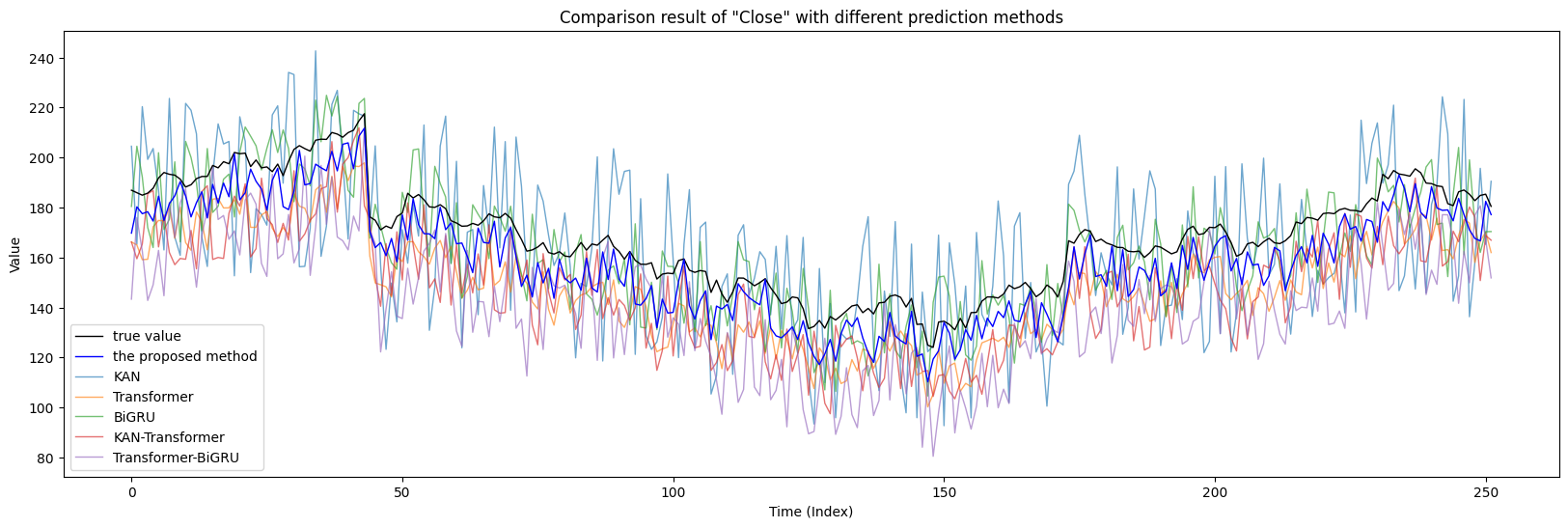}
    \label{fig:short-b}
  \caption{The multi-layer hybrid MTL structure for stock market prediction, involving an enhanced Transformer, KAN and BiGRU.}
  \label{fig:short}
\end{figure*}
\begin{figure*}
  \centering
    %\fbox{\rule{0pt}{2in} \rule{.9\linewidth}{0pt}}
    \includegraphics[width=1\linewidth]{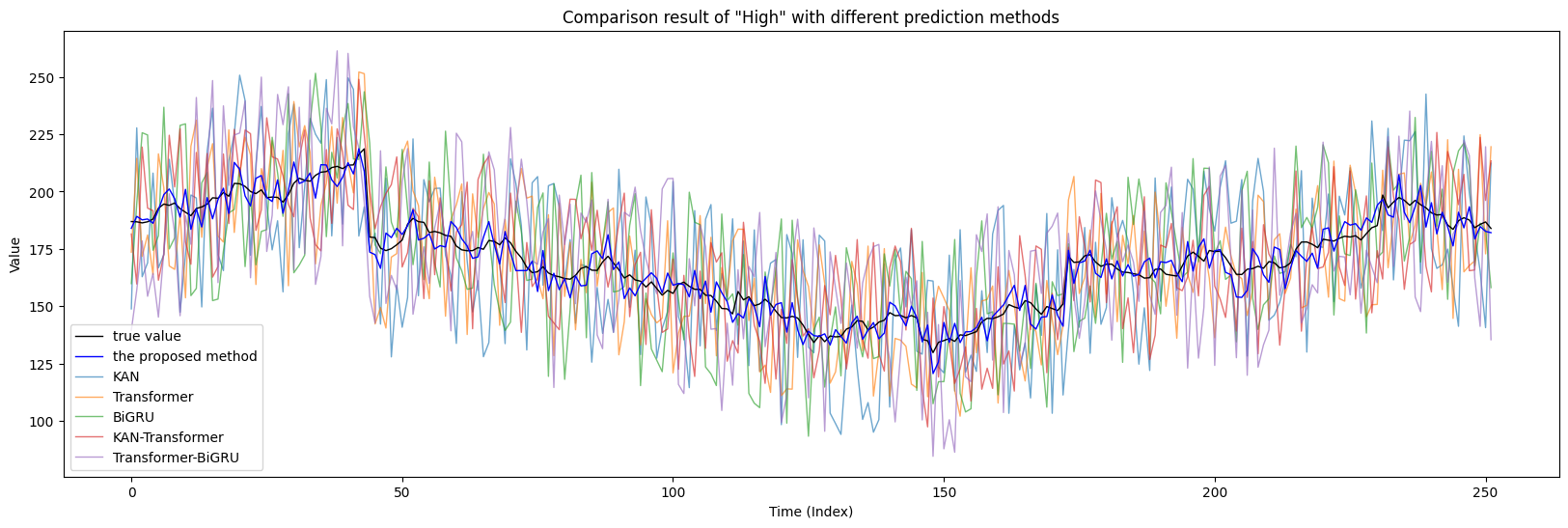}
    \label{fig:short-b}
  \caption{The multi-layer hybrid MTL structure for stock market prediction, involving an enhanced Transformer, KAN and BiGRU.}
  \label{fig:short}
\end{figure*}
\begin{figure*}
  \centering
    %\fbox{\rule{0pt}{2in} \rule{.9\linewidth}{0pt}}
    \includegraphics[width=1\linewidth]{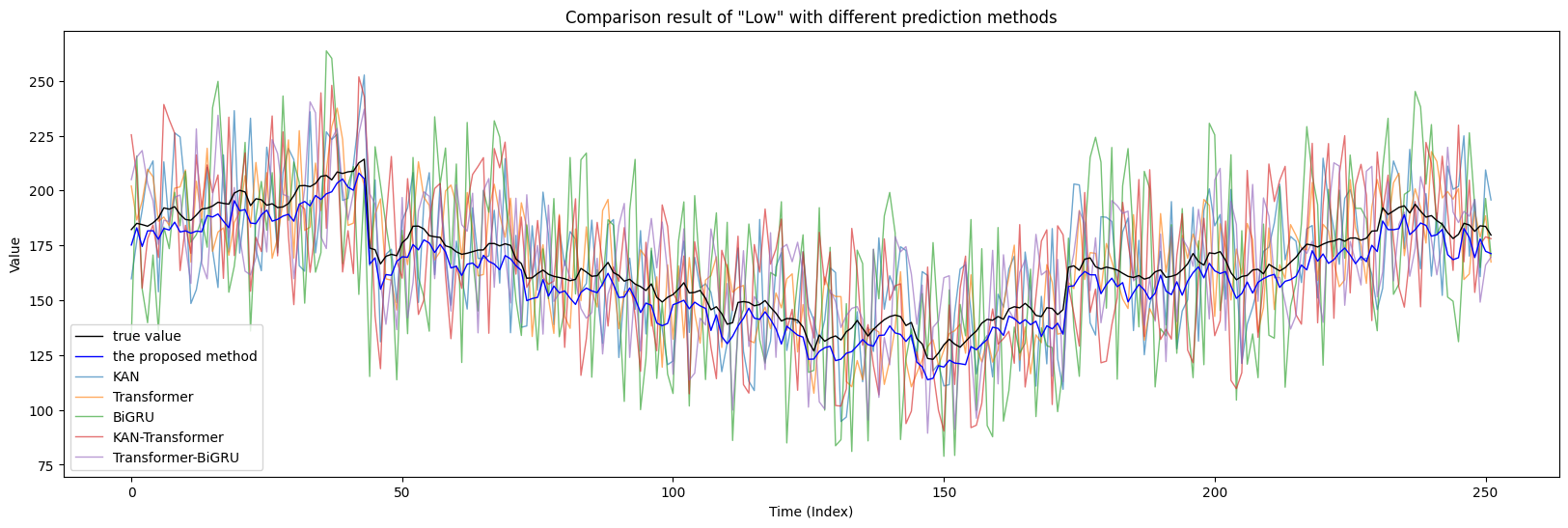}
    \label{fig:short-b}
  \caption{The multi-layer hybrid MTL structure for stock market prediction, involving an enhanced Transformer, KAN and BiGRU.}
  \label{fig:short}
\end{figure*}
\begin{figure*}
  \centering
    %\fbox{\rule{0pt}{2in} \rule{.9\linewidth}{0pt}}
    \includegraphics[width=1\linewidth]{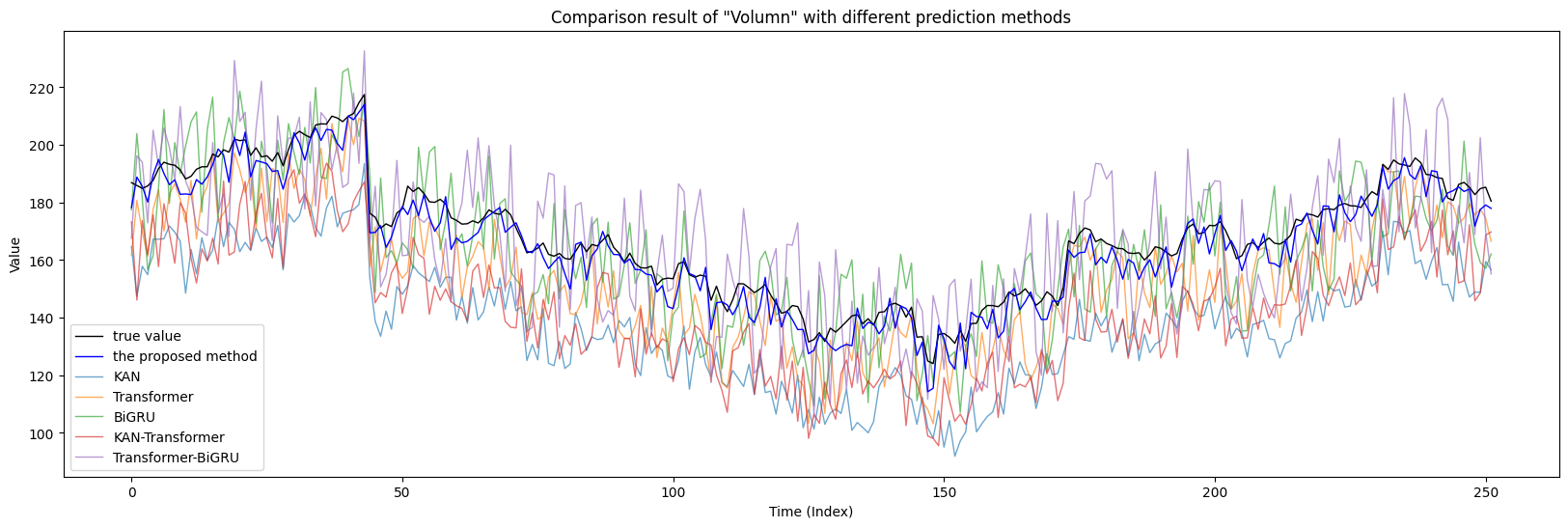}
    \label{fig:short-b}
  \caption{The multi-layer hybrid MTL structure for stock market prediction, involving an enhanced Transformer, KAN and BiGRU.}
  \label{fig:short}
\end{figure*}

Table 2 depicts the average values of the four evaluated metrics from all the comparative advanced methods. It demonstrates that the proposed method achieves significant improvements over all the comparative models. Specifically, the RMSE of the proposed perdition method is 39.820, indicating a 17.2\% reduction compared to Hemajothi et al. \cite{44}, which shows greater robustness in handling large fluctuations and reducing the impact of outliers. Forthermore, the R² reaches 0.977, outperforming Gao et al. \cite{43} and Hemajothi et al. \cite{44} by 4.2\% and 3.1\%, respectively. This highlights the model’s ability to accurately capture both short-term variations and long-term trends, reflecting superior its predictive performance. While improvements are also seen in MAE and MAPE, the gains in RMSE and R² are more critical. These results confirm that the method excels in error control and trend detection, making it well-suited for complex, noisy, and nonlinear time series forecasting.
\begin{table}[htbp]
\centering
\caption{The average values of the four evaluated metrics.}
\begin{tabular}{lllll}
\hline
\textbf{Methods} & \textbf{MAE} & \textbf{RMSE} & \textbf{MAPE} & \textbf{$R^2$} \\
\hline
KAN & 39.278 & 49.683 & 0.056 & 0.916 \\
Transformer & 39.047 & 42.917 & 0.063 & 0.935 \\
BiGRU & 32.741 & 47.050 & 0.045 & 0.940 \\
KAN-Transformer & 29.384 & 39.693 & 0.038 & 0.953 \\
Transformer-BiGRU & 27.046 & 34.821 & 0.051 & 0.950 \\
\textbf{Proposed method} & 18.410 & 21.004 & 0.031 & 0.968 \\
\hline
\end{tabular}
\end{table}

\subsection{ Result Evaluation and Analysis with Ablation Methods}
Fig. 9, Fig.10, Fig.11,Fig 12 as well as Fig. 11 illustrates four representative comparison results of the four stock features with different comparative ablation methods. Compared with models, including KAN, Transformer, BiGRU, KAN-Transformer and Transformer-BiGRU, the proposed method shows significant advantages in prediction accuracy and stability. It can be observed from the figures that in the high-frequency fluctuation region, the prediction results of other models have obvious noise interference, their curve fluctuates greatly, and there is a large deviation from the real value in many points. In contrast, the proposed method is not only smoother, but also highly consistent with the true value at most of the time steps. In terms of peak and trough capture, the proposed method shows strong sensitivity and accuracy. The prediction results of multiple key inflection points such 
as 50-70 and 170-200 time steps, showing that it can accurately track the changes of local highs and lows, while other models have certain prediction lag or over-fitting in these regions. In addition, in the trend recovery interval such as 150-250 time steps, the prediction curves of other models are often amplified by noise and produce large fluctuations. The proposed method can effectively filter out these noises and ensure that the prediction results are more stable.
\begin{figure*}
  \centering
    %\fbox{\rule{0pt}{2in} \rule{.9\linewidth}{0pt}}
    \includegraphics[width=1\linewidth]{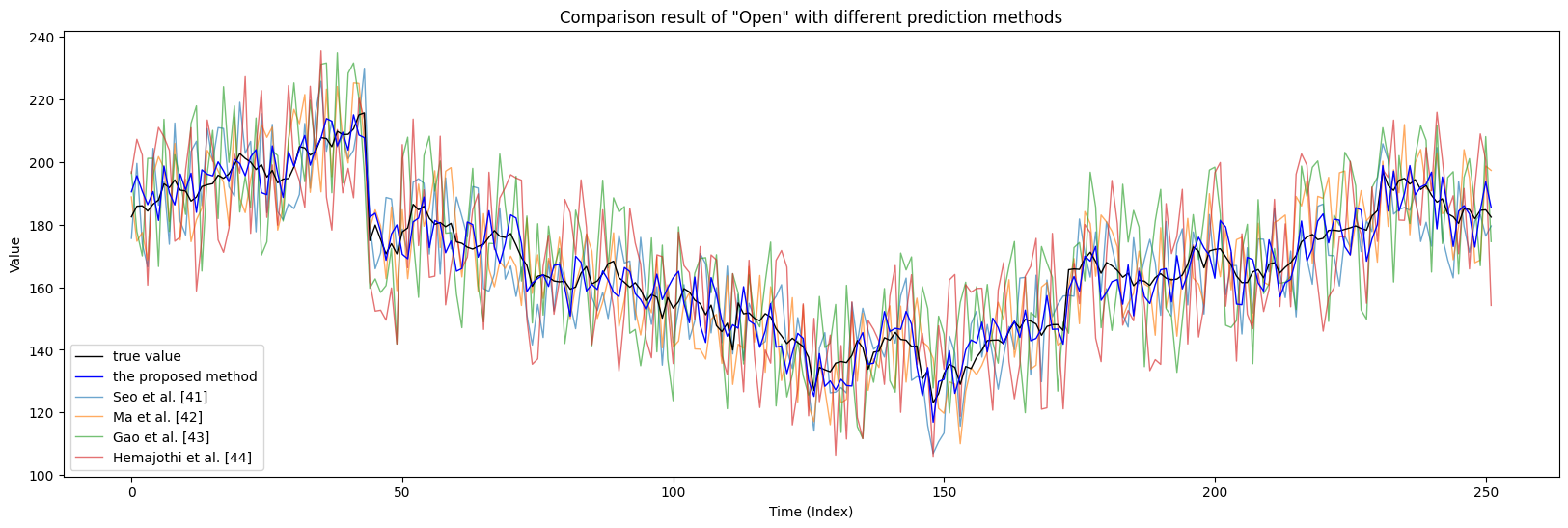}
    \label{fig:short-b}
  \caption{The multi-layer hybrid MTL structure for stock market prediction, involving an enhanced Transformer, KAN and BiGRU.}
  \label{fig:short}
\end{figure*}
\begin{figure*}
  \centering
    %\fbox{\rule{0pt}{2in} \rule{.9\linewidth}{0pt}}
    \includegraphics[width=1\linewidth]{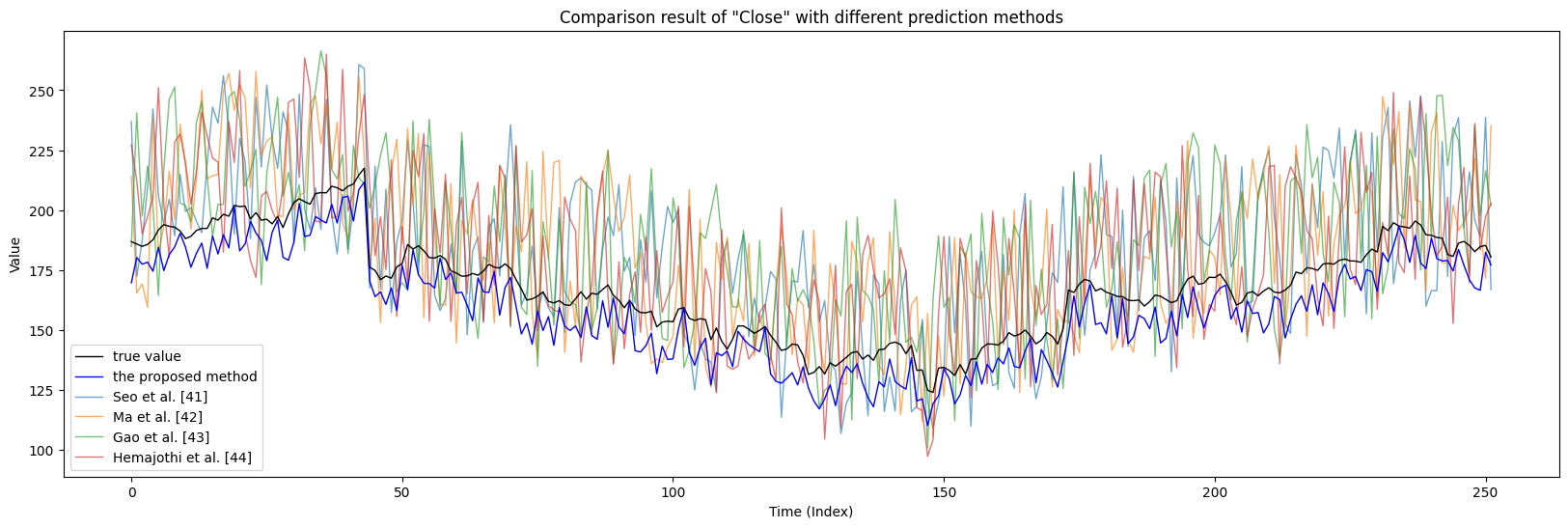}
    \label{fig:short-b}
  \caption{The multi-layer hybrid MTL structure for stock market prediction, involving an enhanced Transformer, KAN and BiGRU.}
  \label{fig:short}
\end{figure*}
\begin{figure*}
  \centering
    %\fbox{\rule{0pt}{2in} \rule{.9\linewidth}{0pt}}
    \includegraphics[width=1\linewidth]{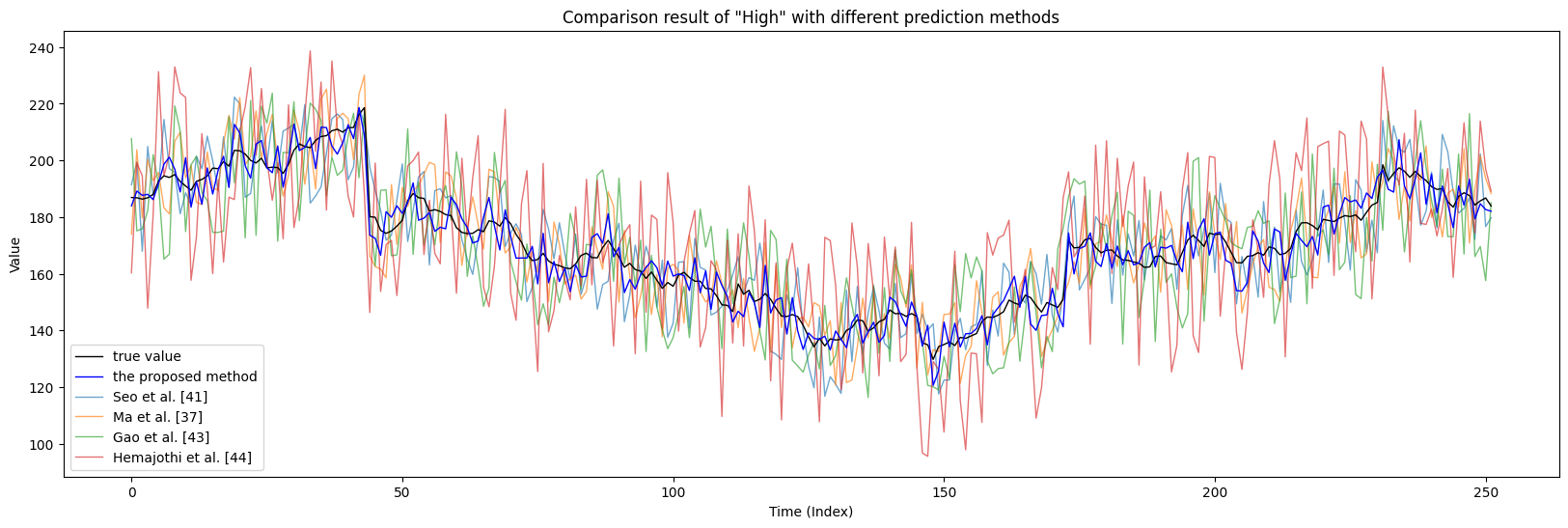}
    \label{fig:short-b}
  \caption{The multi-layer hybrid MTL structure for stock market prediction, involving an enhanced Transformer, KAN and BiGRU.}
  \label{fig:short}
\end{figure*}
\begin{figure*}
  \centering
    %\fbox{\rule{0pt}{2in} \rule{.9\linewidth}{0pt}}
    \includegraphics[width=1\linewidth]{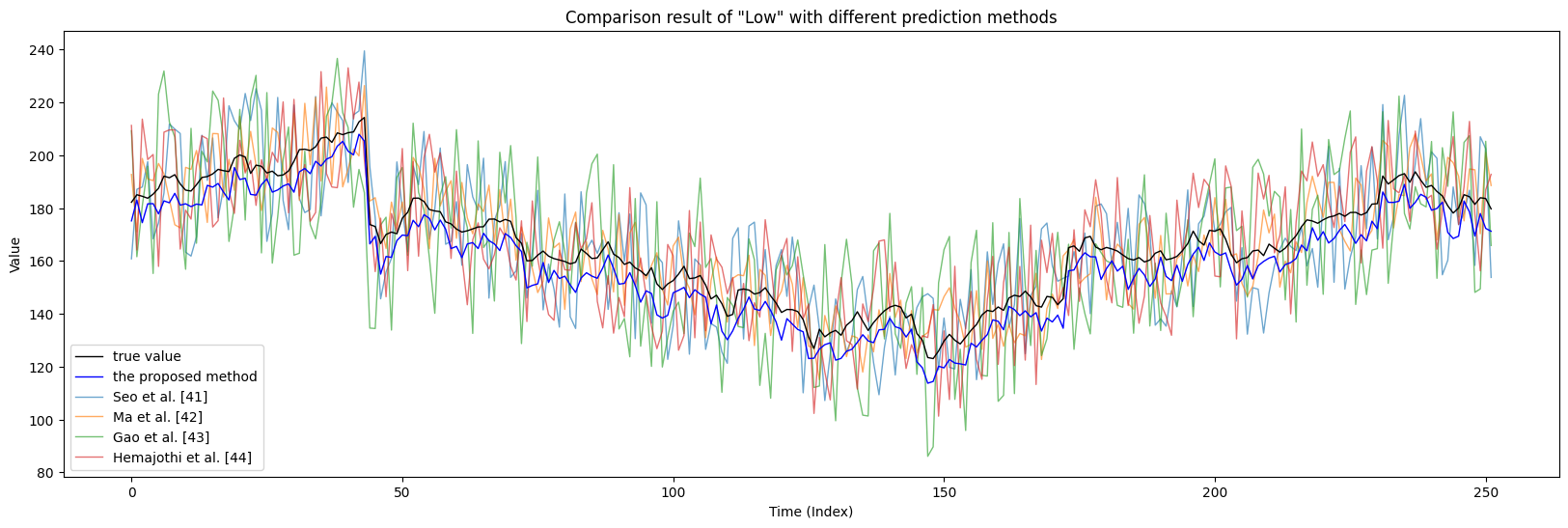}
    \label{fig:short-b}
  \caption{The multi-layer hybrid MTL structure for stock market prediction, involving an enhanced Transformer, KAN and BiGRU.}
  \label{fig:short}
\end{figure*}
\begin{figure*}
  \centering
    %\fbox{\rule{0pt}{2in} \rule{.9\linewidth}{0pt}}
    \includegraphics[width=1\linewidth]{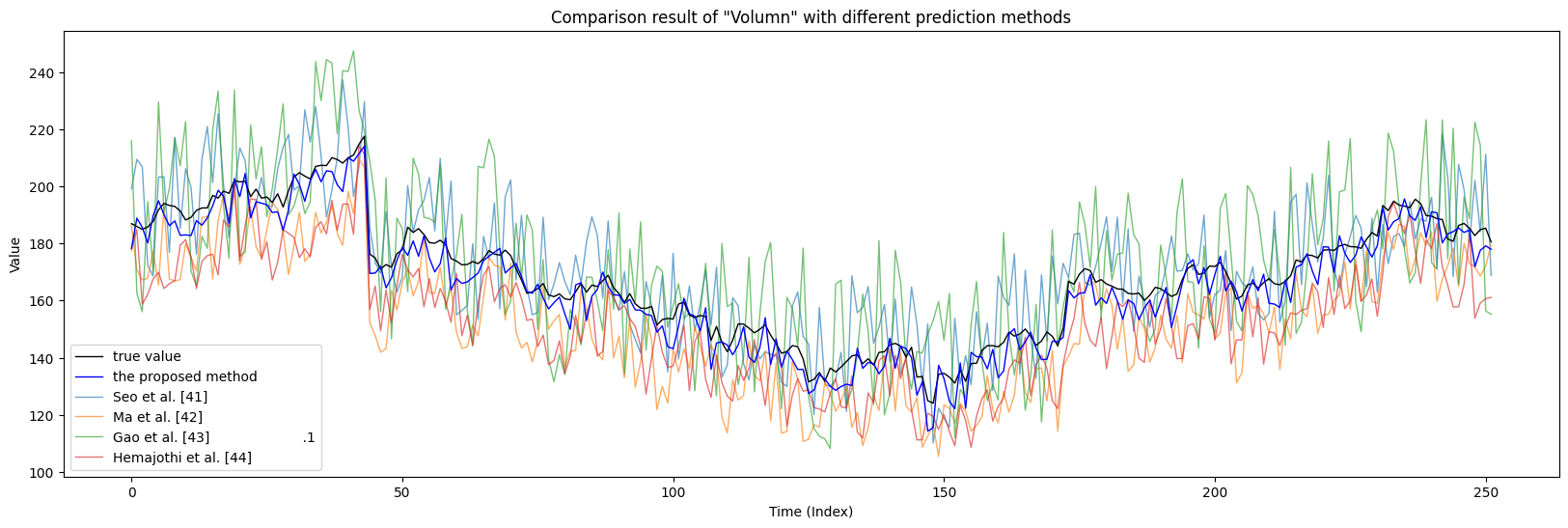}
    \label{fig:short-b}
  \caption{The multi-layer hybrid MTL structure for stock market prediction, involving an enhanced Transformer, KAN and BiGRU.}
  \label{fig:short}
\end{figure*}

Table 3 describes the average values of the four evaluated metrics from all the comparative ablation methods. It shows that the proposed method presents significant advantages in RMSE and R2, highlighting its robustness and trend capture 
ability in complex data environments. The RMSE is 21.004, which is 39.7 \% less than the best performing Transformer-BiGRU, indicating that the model is more stable when dealing with different high volatility data, especially stock data. On R2, the proposed method reaches 0.968, which is significantly better than other comparison models, verifying its stronger fitting ability.  Although MAE and MAPE have also improved, the improvement of RMSE and R2 is particularly important. The performance of these key indicators proves that our model can not only effectively reduce errors, but also accurately capture long-term trends, providing more reliable support for series prediction tasks.
\begin{table}[htbp]
\centering
\caption{The average values of the four evaluated metrics.}
\begin{tabular}{lllll}
\hline
\textbf{Methods} & \textbf{MAE} & \textbf{RMSE }& \textbf{MAPE} & \textbf{$R^2$} \\
\hline
Seo et al. \cite{41} & 22.103 & 58.475 & 0.063 & 0.925 \\
Ma et al. \cite{42} & 19.236 & 52.049 & 0.049 & 0.914 \\
Gao et al. \cite{43} & 24.859 & 51.816 & 0.064 & 0.938 \\
Hemajothi et al.\cite{44} & 20.934 & 48.068 & 0.051 & 0.948 \\
\textbf{Proposed method} & 16.007 & 39.820 & 0.032 & 0.977 \\
\hline
\end{tabular}
\end{table}

The intricate structure and computational processes of Transformer encoders and KAN layers often lead to elevated time complexity. However, our model effectively addresses this limitation, demonstrating a significant improvement in inference efficiency compared to state-of-the-art methods. The detailed comparative results are presented in the table below.

\begin{table}[htbp]
\centering
\caption{The inference time of the four evaluated metrics}
\begin{tabular}{ll}
\hline
\textbf{Methods} & \textbf{Time} \\
\hline
Seo et al. \cite{41} & 0.0284 \\
Ma et al. \cite{42} & 0.0199 \\
Gao et al. \cite{43} & 0.0213 \\
Hemajothi et al. \cite{44} & 0.0263 \\
\textbf{Proposed method} & 0.0193 \\
\hline
\end{tabular}
\end{table}

Furthermore, we performed cross-validation on the experimental results and calculated the average values to ensure the robustness and reliability of our findings. The results of the cross-validation are consistent with the overall performance, with  an Average Test R² of 0.9831.

\section{CONCLUSION}
This study proposed a multi-layer hybrid MTL structure to address the challenges of multi-task stock price prediction caused by the market’s high volatility, complexity, and dynamic nature. The framework novelty integrates an enhanced Transformer encoder for feature extraction, a BiGRU for capturing long-term temporal dependencies, and a KAN to enhance the learning process. Experimental results demonstrate that the proposed learning network outperforms competitive models, achieving MAE as low as 0.45 and $R^2$ up to 0.98, highlighting its robustness and predictive accuracy. These results confirm the effectiveness of leveraging complementary learning techniques to capture complex relationships and improve forecasting performance. The framework offers a promising approach for future research and applications in stock market prediction.

%%%%%%%%% REFERENCES
{\small
\bibliographystyle{unsrt}
\bibliography{egbib}
}

\end{document}